\pdfoutput=1

\documentclass[11pt]{article}

\usepackage{acl}

\usepackage{times}
\usepackage{latexsym}
\usepackage{multirow}
\usepackage{graphicx}
\usepackage{amsmath}
\usepackage{amssymb}
\usepackage{booktabs}
\usepackage{algorithm2e}

\usepackage{subcaption}
\usepackage{caption}
\usepackage{bbding}
\usepackage{colortbl}
\usepackage{tcolorbox}
\usepackage[T1]{fontenc}
\usepackage{xcolor}         
\usepackage{color}
\usepackage{colortbl,array}
\usepackage[rgb]{xcolor}
\usepackage{tikz}
\usetikzlibrary{tikzmark}
\usepackage{arydshln}
\makeatletter

\usepackage[normalem]{ulem}
\useunder{\uline}{\ul}{}
\newcommand*\myfontsize{%
  \@setfontsize\myfontsize{8}{8}%
}

\makeatother

\definecolor{myred}{rgb}{0.8, 0.1, 0.1}
\definecolor{mygray}{rgb}{0.9, 0.9, 0.9}
\definecolor{myblue}{HTML}{054488}
\definecolor{mylightblue}{RGB}{235,245,250}
\definecolor{mygreen}{HTML}{056b34}
\definecolor{myorange}{HTML}{ff8800}
\definecolor{mypurple}{HTML}{8400ff}
\definecolor{mypink}{HTML}{f7acb9}

\definecolor{green}{RGB}{0,120,0}
\definecolor{deepblue}{RGB}{0,0,255}
\definecolor{blue}{RGB}{0,179,241}
\definecolor{orange}{RGB}{200,110,0}
\definecolor{purple}{RGB}{120,0,160}

\newcommand{\ie}{\textit{i.e.}}
\newcommand{\eg}{\textit{e.g.}}
\usepackage{algorithm}
\usepackage{algorithmic}

\usepackage[T1]{fontenc}

\usepackage[utf8]{inputenc}

\usepackage{microtype}
\usepackage{bbold}

\usepackage{inconsolata}

\newcommand{\ours}{\texttt{Agentic-R}}

\title{Agentic-R: Learning to Retrieve for Agentic Search}

\author{Wenhan Liu$^1$, Xinyu Ma$^2$, Yutao Zhu$^1$, Yuchen Li$^2$, \textbf{Daiting Shi}$^2$ \\ 
\textbf{Dawei Yin}$^2$ \and \textbf{Zhicheng Dou}$^{1}$\thanks{Corresponding author.} \\
$^1$Gaoling School of Artificial Intelligence, Renmin University of China \\
$^2$Baidu Inc., Beijing, China \\
\texttt{lwh@ruc.edu.cn, xinyuma2016@gmail.com, dou@ruc.edu.cn}
}

\begin{document}
\maketitle

\begin{abstract}
Agentic search has recently emerged as a powerful paradigm, where an agent interleaves multi-step reasoning with on-demand retrieval to solve complex questions.
Despite its success, how to design a retriever for agentic search remains largely underexplored.
Existing search agents typically rely on similarity-based retrievers, while similar passages are not always useful for final answer generation. 
In this paper, we propose a novel retriever training framework tailored for agentic search.
Unlike retrievers designed for single-turn retrieval-augmented generation (RAG) that only rely on local passage utility, we propose to use both local query-passage relevance and global answer correctness to measure passage utility in a multi-turn agentic search.
We further introduce an iterative training strategy, where the search agent and the retriever are optimized bidirectionally and iteratively. Different from RAG retrievers that are only trained once with fixed questions, our retriever is continuously improved using evolving and higher-quality queries from the agent.
Extensive experiments on seven single-hop and multi-hop QA benchmarks demonstrate that our retriever, termed \ours{}, consistently outperforms strong baselines across different search agents. Our codes are available at: \url{https://github.com/8421BCD/Agentic-R}.

\end{abstract}

\section{Introduction}\label{sec:intro}
\begin{figure}[t]
	\centering
	\includegraphics[width=1\linewidth]{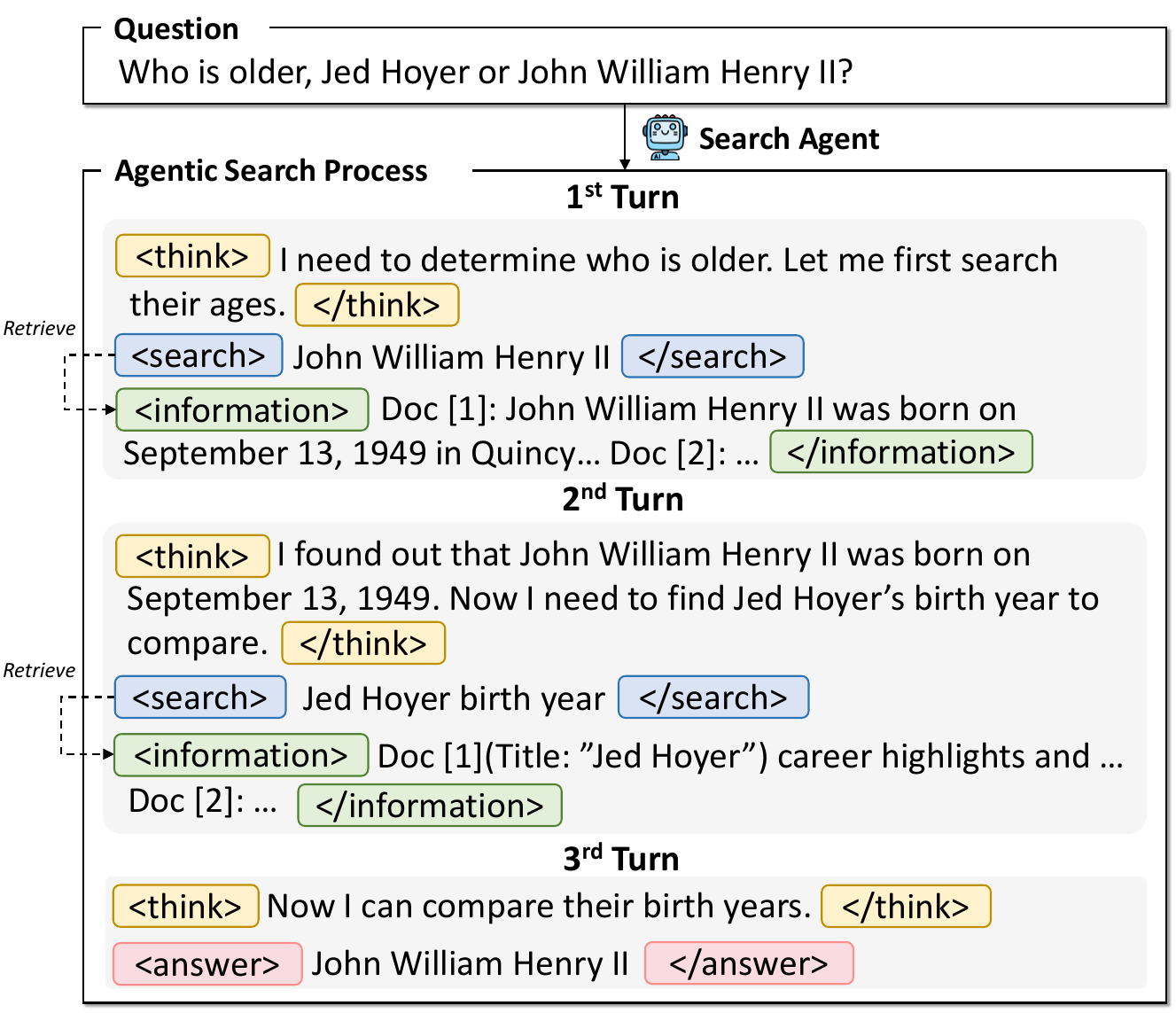}
	\caption{An example of agentic search.}\label{fig:agentic_search}        
\end{figure}

Retrieval-augmented generation (RAG)~\cite{AsaiMZC23,rag_survey,flashrag} has become a widely adopted approach to address the knowledge limitations of large language models (LLMs) by retrieving external information to support generation.
Recently, advances in large reasoning models~\cite{r1} have given rise to a new paradigm known as \emph{agentic search}~\cite{search-r1,search-o1}. This approach extends traditional RAG from a single-turn retrieval to a multi-step ``search-during-think'' process (as shown in Figure~\ref{fig:agentic_search}).
Unlike traditional single-turn RAG, agentic search enables an agent to decompose complex questions into a sequence of sub-queries, interleave reasoning with retrieval across multiple turns, and progressively gather evidence before producing a final answer.
By integrating reasoning with retrieval, agentic search achieves superior performance compared to traditional single-turn RAG on challenging NLP tasks.

Despite this progress, most existing research on agentic search focuses on designing more powerful agents~\cite{autorefine,search-r1}, with less attention paid to optimizing a critical component: \textbf{retriever}.
This is a critical oversight because the quality of retrieved passages directly impacts both the training process and the final inference performance of the agent.
Currently, most approaches employ \textit{off-the-shelf} retrievers, such as E5~\cite{e5} or BGE~\cite{bge}, which rely on semantic similarity.
However, as demonstrated in previous studies~\cite{SCARLet,abs-2404-03302}, high semantic similarity does not guarantee that a passage is useful for generating answer.

To better align retrieval with downstream generation, prior studies have explored modeling passage utility for downstream generation and training utility-optimized retrievers for RAG~\cite{replug,SCARLet,llm-embedder}.
These methods typically estimate utility by feeding retrieved passages into a generator and measuring downstream performance, such as generation likelihood or task-specific metrics.
While effective for \emph{single-turn} RAG, extending these approaches to \emph{multi-turn} agentic search presents significant challenges:
\textbf{First}, utility modeling in single-turn RAG relies on gold answers, which are available in existing datasets, to evaluate whether a passage addresses the question. In contrast, agentic search involves intermediate queries generated by the agent in each turn, but gold answers for these intermediate queries are not available, making it challenging to evaluate passage utility based on the gold answer. \textbf{Second}, passage utility in agentic search goes beyond local relevance. A passage relevant to current sub-query may contain misleading information that steers the subsequent reasoning process in the wrong direction, which ultimately leads to an incorrect final answer~\cite{worse_than_zero_shot,power_of_noise,rethinking_relevance}. Thus, relying solely on local relevance to evaluate passage utility is insufficient.

Furthermore, existing retriever training methods for RAG~\cite{replug,SCARLet} typically adopt a \emph{one-way optimization} paradigm, where the retriever is trained only once based on fixed training queries (\ie, user questions) and utility signals provided by a fixed generator. However, this approach is sub-optimal for training retrievers for agentic search.
Unlike traditional RAG, the training queries for retrievers in agentic search are generated by the search agent itself.
After the retriever is optimized, the search agent can further improve through reinforcement learning by interacting with this stronger retriever~\cite{Empirical_Study}.
The improved search agent could generate new search trajectories with higher-quality queries, which can be leveraged to further optimize the retriever.
Therefore, the optimization of the retriever and the search agent should be formulated as a \emph{bidirectional and iterative} process.



In this paper, we propose the first retriever training framework designed specifically for agentic search.
We introduce a passage utility modeling strategy that considers both the single-turn relevance and the correctness of the final answer.
To evaluate single-turn relevance, we design an LLM-based listwise scoring approach that generates the relevance scores for multiple candidate passages of each intermediate query. 
To measure the correctness of the final answer, we evaluate whether the agent can derive the correct final answer when using a specific passage.
Furthermore, we propose an iterative agent--retriever optimization framework that alternates between training the search agent and the retriever, allowing them to evolve together and finally resulting in a stronger retriever for agentic search. 
We conduct extensive experiments on seven benchmarks, including both multi-hop and single-hop question answering datasets. The results demonstrate that our method consistently outperforms strong baselines across different search agents.
Further analysis shows that our \ours{} also make the agent solve questions with fewer search queries.

Our contributions are threefold:

$\bullet$~We present the first retriever training framework specifically designed for search agents, addressing a critical yet underexplored component in existing agentic search systems.

$\bullet$~Based on the multi-turn nature of agentic search, we propose a novel passage utility modeling approach that considers both the relevance of the passage to current search query and its contribution to the correctness of the final answer.

$\bullet$~We introduce an iterative training framework, where the search agent and the retriever are optimized bidirectionally to progressively improve the retriever.



\section{Related Work}
\paragraph{Agentic Search}
Retrieval-Augmented Generation (RAG) significantly enhances Large Language Models (LLMs) by incorporating external knowledge sources~\cite{AsaiMZC23, rag_survey, RamLDMSLS23}. A fundamental challenge in RAG systems involves determining the optimal timing and approach for retrieval~\cite{ShaoGSHDC23, TrivediBKS23, ManuSearch}. Previous studies explored prompt-based approaches that enable interleaved reasoning and retrieval~\cite{ircot, ReAct} and supervised fine-tuning (SFT) approaches that learn to call the search engine~\cite{self-rag, Toolformer}. Recently, reinforcement learning (RL) has emerged as a powerful and scalable alternative, enabling agents to learn complex, multi-turn search strategies directly from task-outcome rewards without relying on extensive supervision~\cite{search-r1, r1-searcher, ReSearch, DeepResearcher}. This paradigm allows LLMs to dynamically decompose questions, formulate sequential queries, and retrieve information across turns. Despite significant progress in optimizing the search agents, the retriever, which is another critical component and significantly influences the agent's performance~\cite{Empirical_Study}, remains largely underexplored. In this paper, we propose to train a retriever tailored for agentic search.

\paragraph{Training Retrievers for Generation}
A key challenge in RAG systems is the gap between retrieval objectives (topical similarity) and generation needs (passage utility). To align retrievers with downstream tasks, recent work focuses on training utility-oriented retrievers using feedback from the generator. To obtain the passage utility, existing studies explore various approaches, such as using the generation likelihood of ground-truth answers~\cite{replug, SCARLet, atlas}, downstream task metrics~\cite{stochastic-rag, SimLM}, and LLM-based annotation~\cite{llm_utility}. 
However, such passage utility modeling is limited to single-turn RAG, and the corresponding retriever training follows a one-way optimization paradigm which is sub-optimal for agentic search.
In this paper, we propose a passage utility modeling mechanism and an agent--retriever iterative optimization framework tailored for training retrievers in agentic search.


\section{Preliminary: Agentic Search}
We consider a search agent based on a Large Language Model (LLM) that alternates between reasoning and external retrieval over multiple turns. An example is shown in Figure~\ref{fig:agentic_search}.
In each turn $i$, the agent first produces a reasoning trace $t_i$, to analyze the current context and assess what information is still missing.
Conditioned on this reasoning, the agent generates a search query $q_i$.
After that, the retriever returns a set of passages $D_i$ that are incorporated into the agent’s context for subsequent reasoning.

This reasoning-retrieval cycle repeats across iterations, allowing the agent to progressively refine its understanding and gather relevant information.
Once the agent determines that the accumulated information is sufficient, it terminates the retrieval process and produces the final answer.
During the whole process, the reasoning steps, search query, retrieved passages and the final answer are explicitly enclosed within tags, such as \texttt{<think>} \texttt{</think>}, \texttt{<search>} \texttt{</search>}, \texttt{<information>} \texttt{</information>} and \texttt{<answer>} \texttt{</answer>}, respectively.

\begin{figure*}[!tb]
  \centering
  \includegraphics[width=1\linewidth]{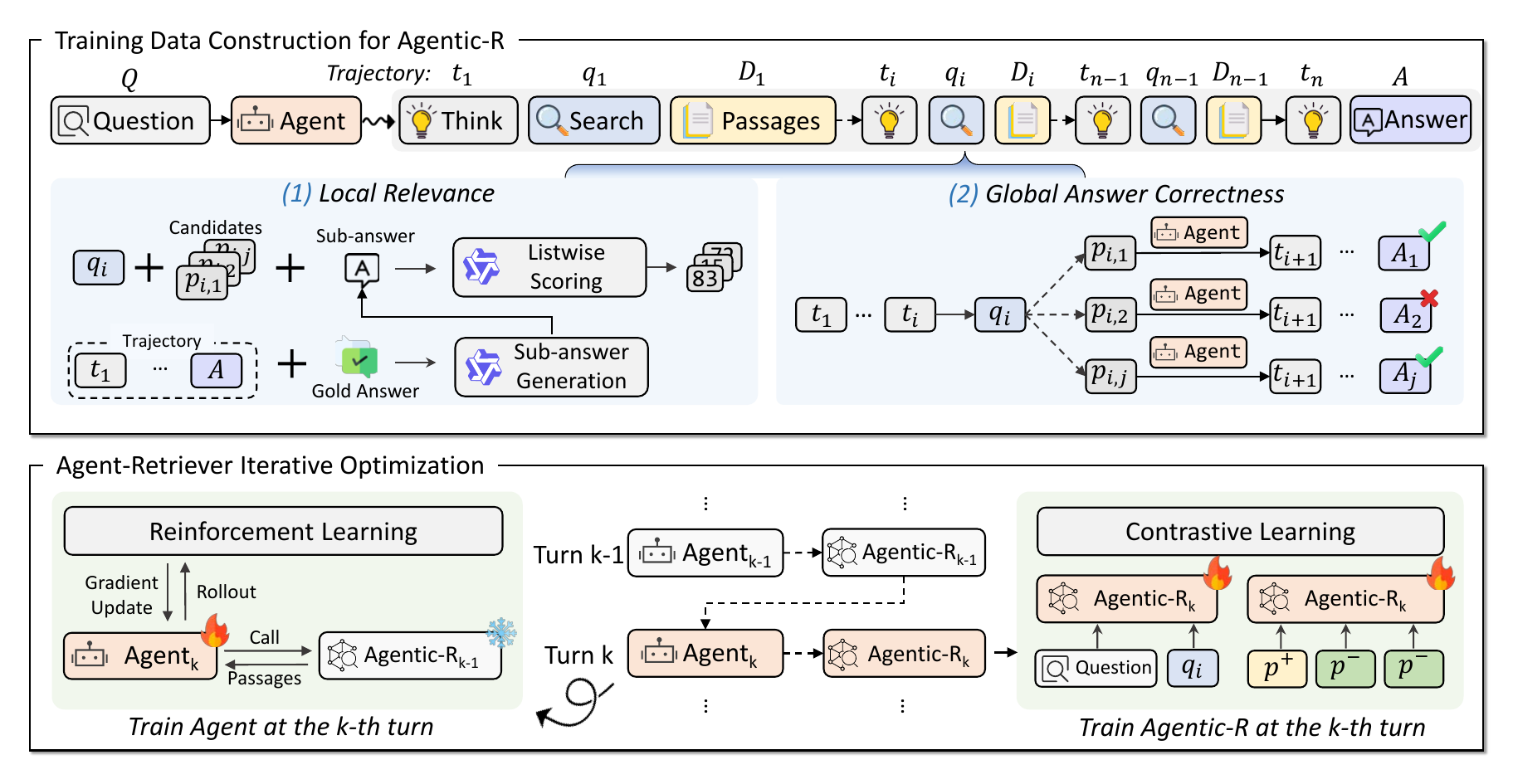}
  \caption{An overview of our training framework.}\label{fig:model}  
\end{figure*}

\section{Methodology}
\subsection{\ours} \label{sec:agentic_retriever}
In this section, we first describe how we construct training data for \ours{} by modeling passage utility in agentic search, and then introduce the corresponding training approach.

\subsubsection{Training Data Construction}
\label{sec:utility_modeling}
In this section, we propose to measure passage utility in agentic search from both local and global perspectives, and to use this utility to distinguish positive and negative passages for subsequent retriever training.
Given an original user question $Q$ and a retriever $\mathcal{R}$, we first let the search agent generate the whole trajectory,
denoted as $\mathcal{T} = \{t_1, q_1, D_1, \dots, t_i, q_i, D_i, \dots, t_{n}, A\}$,
where $t_i$, $q_i$ and $D_i$ represent the reasoning content, search query and top passages returned by retriever $\mathcal{R}$ at the $i$-th turn, and $A$ is the final generated answer. For each query $q_i$, we first retrieve a candidate passage set
$\mathcal{P}_i = \{p_{i,1}, \dots, p_{i,j}, \dots\}$ from the passage corpus using retriever $\mathcal{R}$, where $|\mathcal{P}_i| = 20$.
Then, we propose to measure the utility of each candidate passage $p_{i,j}$ from two different perspectives: (1) local relevance and (2) global answer correctness, based on which we construct the positive and negative passages.

\paragraph{Local Relevance.}
Local relevance measures whether a candidate passage $p_{i,j}$ can answer the query $q_i$ in $i$-th turn.
Unlike prior single-turn RAG methods~\cite{replug,stochastic-rag} that rely on gold answers to measure the relevance, we do not have gold answers for queries of each turn.
In this part, we design an LLM-based listwise scoring approach to evaluate the local query-passage relevance.

Specifically, we input the $q_i$ and all its candidate passages $\mathcal{P}_i$ into a strong LLM, Qwen2.5-72B-Instruct\footnote{\url{https://huggingface.co/Qwen/Qwen2.5-72B-Instruct}} and instruct the LLM to assign a relevance score in the range $[0, 100]$ to each passage $p_{i,j}$, where higher scores indicate stronger relevance. We divide the score range into five intervals with fine-grained scoring rules. For example, scores between 81 and 100 indicate that the passage directly addresses the search query or explicitly contains the required answer. The design of listwise scoring is inspired by listwise passage reranking in Information Retrieval~\cite{rankgpt,fullrank,reasonrank, coranking}. Different from pointwise relevance scoring, which evaluates a single passage, listwise scoring compares multiple passages simultaneously and could yield more accurate relevance scores.

To help LLM better assess the relevance, besides $q_i$ and $p_{i,j}$, we also incorporate a sub-answer into the relevance scoring prompt (shown in Figure~\ref{fig:scoring_with_answer}). The sub-answer is generated by prompting the same LLM with the full search trajectory $\mathcal{T}$ and the final gold answer, and asking it to infer a correct sub-answer corresponding to each intermediate query $q_i$. If the LLM is unable to infer a reliable sub-answer, it will output ``not sure'', in which case no sub-answer will be provided in the listwise scoring prompt (prompt shown in Figure~\ref{fig:scoring_without_answer}). The prompt for generating the sub-answer is shown in Figure~\ref{fig:subanswer_generation}. Formally, we define the local relevance score $\mathrm{LR}_{i,j}$ for passage $p_{i,j}$ in $i$-th search turn as:
\begin{equation}
\label{equ:lr}
\{\mathrm{LR}_{i,1}, \ldots, \mathrm{LR}_{i,j}\}
=
\text{LLM}\!\left(
q_i,\; \mathcal{P}_i,\; A_i^{\text{sub}}
\right),
\end{equation}
where $\mathcal{P}_i = \{p_{i,1}, \ldots, p_{i,j}\}$ denotes the candidate passages for query $q_i$ and $A_i^{\text{sub}}$ is an optional sub-answer for $q_i$.
If no reliable sub-answer can be inferred, we set $A_i^{\text{sub}} = \varnothing$.

\paragraph{Final Answer Correctness.}
As discussed in Section~\ref{sec:intro}, passage with high local relevance does not guarantee that it could lead the search agent to generate a correct final answer. In this part, we explicitly incorporate final answer correctness as another passage utility.

Given query $q_i$, we concatenate each passage $p_{i,j}$ with $q_i$ and let the agent continue generation conditioned on $p_{i,j}$ together with the preceding search trajectory, until a final answer $A_{i,j}$ is produced.
\begin{equation}
\label{eq:final_answer_generation}
\{t_{i+1}, \dots, A_{i,j}\}
=
\mathrm{Agent}\!\left(\{t_1, \dots, q_i, p_{i,j}\}\right).
\end{equation}

After that, we compute the exact match (EM) metric between the generated answer $A_{i,j}$ and the gold answer $A^{\text{gold}}$ as the global answer correctness $\mathrm{GAC}_{i,j}$:
\begin{equation}
\label{equ:gac}
\mathrm{GAC}_{i,j} = \mathrm{EM}(A_{i,j}, A^{\text{gold}}).
\end{equation}
Although the final answer is influenced not only by $p_{i,j}$ but also by subsequent turns, all candidate passages $p_{i,j}$ are evaluated under the same preceding trajectory and search agent. Therefore, differences in final answer correctness can be a fair and valid metric to measure how $p_{i,j}$ steers the subsequent reasoning and generate the final answer.

\paragraph{Positive and Negative Passages Selection.}
\label{sec:pos_neg}
After computing the local relevance $\mathrm{LR}_{i,j}$ (defined in Eq.~\ref{equ:lr}) and global answer correctness $\mathrm{GAC}_{i,j}$ (defined in Eq.~\ref{equ:gac}) for each candidate passage $p_{i,j}$ of query $q_i$, we rank all candidates based on two sorting keys. Specifically, passages are first ranked by global answer correctness GAC (the first key) in descending order, and passages with identical GAC are further ordered by LR (the second key) in descending order.
We prioritize global answer correctness since it directly reflects final task success: for example, a passage that leads to a correct final answer should be preferred over one that does not, regardless of their local relevance.


After ranking, we select the top-ranked passage as the positive passage, while passages ranked below are sampled as negatives.
The total number of passages (positive plus negatives) is fixed to $N = 16$.
To ensure positive quality, we require its $\mathrm{GAC}_{i,j}=1$ and $\mathrm{LR}_{i,j} \ge 60$.
If no passage satisfies both conditions, the training instance for query $q_i$ is discarded. For all $q_i$ in trajectory $\mathcal{T}$, we use the same method to construct training data.

\subsubsection{Training Approach}
\label{sec:training}

After constructing positive and negative passages, we use contrastive learning~\cite{dpr} to train our retriever model \ours.

\paragraph{Training Input.}
When modeling passage utility, we consider not only the relevance of a passage to the current query $q_i$, but also its contribution to answering the original question $Q$ and generating the final answer.
Thus, besides $q_i$, we also incorporate the original question $Q$ as auxiliary information and concatenate them together as the input $x_i$ to the query encoder:
\begin{equation}
\label{equ:query_input}
x_i = Q \; [\text{SEP}] \; q_i,
\end{equation}
where $[\text{SEP}]$ denotes a separator token. Note that we do not include queries from previous turns as input.
This is because, in agentic search, agent queries are typically self-contained and do not involve anaphoric references (\eg, terms such as ``it'') that require contextual disambiguation. This differs from conversational search~\cite{convsearch} or session search~\cite{WangD023} tasks, where previous queries are necessary to understand the current query. Empirically, we find that incorporating previous queries introduces retrieval noise and degrades final performance. Detailed analysis is provided in Appendix~\ref{app:input_analysis}.

\paragraph{Training Loss.}
In addition to the sampled negatives described in Section~\ref{sec:pos_neg}, we also incorporate in-batch negatives and cross-device negatives~\cite{RocketQA, llm-embedder} to expand the scale of negative passages. Consequently, our method yields $(B \times G \times N - 1)$ negative samples in total, where $B$ is the batch size, $G$ is the number of GPU devices, and $N$ is the number of constructed samples per query. For each training instance $x_i$, the contrastive learning loss is defined as:
\begin{equation}
\mathcal{L}
=
- \log
\frac{\exp\big(\mathrm{sim}(x_i, z^+)\big)}
{\sum_{z \in \mathcal{Z}} \exp\big(\mathrm{sim}(x_i, z)\big)},
\end{equation}
where $z^+$ denotes the positive passage embedding, $\mathcal{Z}$ includes the sampled positives and negatives as well as in-batch and cross-device negatives, and $\mathrm{sim}(\cdot,\cdot)$ denotes the embedding similarity.

\subsection{Agent-Retriever Iterative Optimization}
\label{sec:iterative_optimization}
As discussed in Section~\ref{sec:intro}, the optimized \ours{} could further be used to improve the search agent by providing more relevant passages during the RL process and the improved search agent could generate new training queries to further train the retriever, which is a bidirectional and iterative process. Motivated by this, we propose an iterative training framework that iteratively optimizes the search agent and \ours.
We first describe the training of the search agent, followed by the iterative optimization procedure.


\paragraph{Agent Training.}
We apply the same RL training approach as Search-R1~\cite{search-r1} for our search agent training. Specifically, we adopt the RL algorithm PPO~\cite{ppo} to train the agent. During RL training, the agent generates trajectories by performing multiple turns of interactions with the retriever until generating the final answer. Then, we use exact match (EM) between the generated answer and the gold answer as the final reward. 
The prompt we used for training is shown in Figure~\ref{fig:searchr1_prompt}. Additional training details are provided in Appendix~\ref{app:agent_training}.

\begin{algorithm}[t]
\caption{Iterative Optimization Process}
\label{alg:iterative_optimization}
\KwIn{Training questions $\mathcal{Q}$; initial retriever \ours{$_{0}$}; number of iterations $K$}
\KwOut{Optimized $\mathrm{Agent}_K$ and \ours{$_K$}}

\For{$i = 1$ \KwTo $K$}{
    \textbf{Agent Training:} \\
    Train the search agent $\mathrm{Agent}_i$ using PPO by interacting with retriever \ours{$_{i-1}$}\;

    \textbf{Retriever Training:} \\
    (1) Use $\mathrm{Agent}_i$ to generate trajectories of $\mathcal{Q}$.\; \\
    (2) Retrieve candidate passages using $\mathcal{R}_{i-1}$ and construct positive and negative passage following Section~\ref{sec:utility_modeling}.\; \\
    (3) Train \ours{$_i$} following Section~\ref{sec:training}.
}

\Return{$\mathrm{Agent}_K$, \ours{$_K$}}
\end{algorithm}

\paragraph{Iterative Optimization.}
We adopt an iterative optimization mechanism to iteratively train the search agent and the retriever.
At iteration $i$, we first optimize the search agent $\mathrm{Agent}_i$, based on our retriever from the previous iteration \ours{$_{i-1}$}. Note that for the first iteration, \ours{$_0$} is initialized by embedding model E5. During agent's training, the retriever is kept fixed and treated as part of the RL environment. After training $\mathrm{Agent}_i$, we use it to generate training queries, retrieve candidate passages using \ours{$_{i-1}$} and construct positive and negative passages based on the passage utility modeling described in Section~\ref{sec:utility_modeling}, which will be used to train \ours{$_{i}$}.
The overall procedure is summarized in Algorithm~\ref{alg:iterative_optimization}.

\section{Experiment}
\subsection{Settings}
\paragraph{Evaluation Datasets.}
We conduct experiments on seven question answering (QA) benchmarks covering both multi-hop and single-hop datasets.
For multi-hop QA, we evaluate on HotpotQA~\cite{hotpotqa}, 2WikiMultihopQA (2Wiki)~\cite{2wiki}, Musique~\cite{musique}, and Bamboogle~\cite{bamboogle}.
The single-hop QA datasets include Natural Questions (NQ)~\cite{NQ}, TriviaQA~\cite{triviaqa}, and PopQA~\cite{popqa}.
Exact match (EM) is used as the evaluation metric.

\paragraph{Baselines.}
We compare our approach with two categories of retriever baselines: 

\textbf{General-purpose embedding models.}
These retrievers are trained as universal text encoders to support a broad range of retrieval tasks and are widely adopted in agentic search systems.
We include BGE~\cite{bge} and E5~\cite{e5} as representative baselines, both of which have demonstrated strong performance on standard retrieval benchmarks and are widely used as off-the-shelf retrievers in agentic search.

\textbf{RAG-specific retrievers.}
These methods optimize retrievers using feedback from downstream generation in single-turn RAG settings.
We compare against LLM-Embedder~\cite{llm-embedder}, SCARLet~\cite{SCARLet}, and REPLUG~\cite{replug}, which leverage generation likelihood or task-level signals to train utility-aware retrievers. Additional details of the baseline retrievers are provided in Appendix~\ref{app:baselines}.

\paragraph{Implementation Details.}
Throughout the iterative training process, we construct training data using the training splits of TriviaQA and HotpotQA. For agent training, we initialize the search agent with Qwen2.5-7B-Base. During training and inference, the maximum search turns $n$ and passage retrieval number $m$ of our agent are set as 5 and 3, respectively. For retriever training, we initialize our \ours{} with E5\footnote{\url{https://huggingface.co/intfloat/e5-base-v2}}.
We use the December 2018 Wikipedia dump~\cite{wiki18} as the retrieval corpus for all experiments. We set the iteration number $K$ of agent--retriever optimization as 2, as additional iterations do not yield further improvements (detailed analysis is provided in Section~\ref{sec:iteration_number}). Additional implementation details are provided in Appendix~\ref{app:implementation}.

\begin{table*}[!t]
\small
\centering
\setlength{\tabcolsep}{2.8mm}{
\begin{tabular}{lcccccccc}
\toprule 
\textbf{} & \multicolumn{4}{c}{\textbf{Multi-Hop QA}} & \multicolumn{3}{c}{\textbf{General QA}} & \textbf{} \\
\cmidrule(lr){2-5} \cmidrule(lr){6-8}
\textbf{Methods} & \textbf{HotpotQA} & \textbf{2Wiki} & \textbf{Musique} & \textbf{Bamboogle} & \textbf{NQ} & \textbf{TriviaQA} & \textbf{PopQA} & \textbf{Avg.} \\ \midrule
\rowcolor{mygray}
\multicolumn{9}{c}{\textbf{Our Search Agent (in-domain)}} \\ \midrule
LLM-Embedder & 39.35 & 39.75 & 17.33 & 36.00 & 41.32 & 62.33 & {\ul 42.69} & 39.82 \\
BGE & 41.71 & 39.30 & 16.21 & 38.40 & 40.36 & 63.66 & 41.92 & 40.22 \\
SCARLet & 42.34 & 39.35 & 16.38 & 40.80 & 40.60 & 63.46 & 42.04 & 40.71 \\
E5 & 41.68 & 40.02 & 17.74 & {\ul 44.00} & {\ul 42.18} & 64.72 & 41.82 & 41.74 \\
REPLUG & {\ul 42.63} & {\ul 40.23} & {\ul 18.90} & 41.60 & 41.46 & {\ul 65.78} & 41.85 & {\ul 41.78} \\
\rowcolor{mylightblue} \ours{} & \textbf{45.82} & \textbf{45.30} & \textbf{20.27} & \textbf{48.00} & \textbf{42.43} & \textbf{69.02} & \textbf{44.14} & \textbf{45.00} \\ \midrule
\rowcolor{mygray}
\multicolumn{9}{c}{\textbf{R1-Searcher (out-of-domain)}} \\ \midrule
LLM-Embedder & 41.39 & 45.72 & 18.36 & 33.60 & 38.80 & 56.29 & 40.05 & 39.17 \\
BGE & {\ul 44.36} & 45.86 & 18.36 & 34.40 & 36.34 & 57.19 & 38.95 & 39.35 \\
SCARLet & 44.11 & 45.94 & 18.53 & 35.19 & 36.34 & 57.64 & 38.73 & 39.50 \\
E5 & 43.56 & {\ul 46.33} & {\ul 21.39} & \textbf{44.00} & 39.39 & 58.69 & 38.31 & {\ul 41.67} \\
REPLUG & 40.68 & 39.66 & 18.32 & 41.60 & \textbf{40.94} & {\ul 62.39} & {\ul 42.11} & 40.81 \\
\rowcolor{mylightblue} \ours{} & \textbf{47.68} & \textbf{49.07} & \textbf{22.54} & {\ul 41.60} & {\ul 39.63} & \textbf{62.52} & \textbf{42.43} & \textbf{43.64} \\ \midrule
\rowcolor{mygray}
\multicolumn{9}{c}{\textbf{SimpleDeepSearcher (out-of-domain)}} \\ \midrule
LLM-Embedder & 35.01 & 32.52 & 13.90 & 40.80 & 33.62 & 59.59 & {\ul 37.72} & 36.17 \\
BGE & 37.31 & 33.95 & 14.06 & \textbf{42.40} & 32.82 & 60.73 & 36.83 & 36.87 \\
SCARLet & 37.39 & 33.44 & 14.39 & 35.19 & 32.38 & 60.81 & 36.93 & 35.79 \\
E5 & 36.61 & 34.01 & 15.55 & 40.80 & 33.65 & 62.48 & 37.09 & 37.17 \\
REPLUG & {\ul 37.39} & {\ul 34.08} & {\ul 15.80} & 41.60 & {\ul 33.71} & {\ul 63.59} & 37.68 & {\ul 37.69} \\
\rowcolor{mylightblue} \ours{} & \textbf{39.75} & \textbf{38.04} & \textbf{17.29} & {\ul 41.60} & \textbf{34.62} & \textbf{65.61} & \textbf{39.09} & \textbf{39.43} \\ \bottomrule
\end{tabular}}
\caption{The performance of retrievers on our trained search agent (in-domain) and two other search agents (out-of-domain). Both our search agent and \ours{} are trained after two iterations. The top two rerankers are highlighted in \textbf{bold} and \underline{underlined}.}
\label{tab:main_results}
\end{table*}

\subsection{Overall Performance} \label{sec:overall}
We evaluate \ours{}'s performance on three search agents: our trained search agent, as well as two other search agents, R1-Searcher~\cite{r1-searcher} and SimpleDeepSearcher~\cite{SimpleDeepSearcher}, to evaluate the generalization across different search agents.
The results are reported in Table~\ref{tab:main_results}. From the results, we have the following observations:

\textbf{(1) \ours{} consistently achieves the best average EM score across all three search agents}. \ours{} outperforms the second-best baseline by about 3.2 points on our trained search agent and by roughly 2 points on R1-Searcher and SimpleDeepSearcher. This demonstrates that \ours{} not only performs well on our in-domain search agent, but also generalizes effectively to other search agents.

\textbf{(2) \ours{} yields larger improvements on multi-hop QA than on single-hop QA}. For example, based on our search agent, the average performance gap between \ours{} and REPLUG on multi-hop QA datasets is about 3 points, higher than the 2 points on single-hop QA datasets. This indicates that \ours{} is particularly effective for search agents in multi-hop scenarios.

\textbf{(3) RAG-specific retrievers do not consistently outperform general-purpose retrievers in agentic search.} For example, LLM-Embedder and SCARLet are often inferior to E5 across all three agents. This may be because: (1) the passage utility derived from single-turn RAGs is not directly applicable to multi-turn agentic search; (2) RAG-specific retrievers are trained on user questions instead of agent-generated queries, resulting in a distribution gap of training queries.

\begin{table*}[!t]
\small
\centering
\vspace{1mm}
\setlength{\tabcolsep}{2.3mm}{
\begin{tabular}{lcccccccc}
\toprule
\textbf{} & \multicolumn{4}{c}{\textbf{Multi-Hop QA}} & \multicolumn{3}{c}{\textbf{General QA}} & \textbf{} \\
\cmidrule(lr){2-5} \cmidrule(lr){6-8}
\textbf{Methods} & \textbf{HotpotQA} & \textbf{2Wiki} & \textbf{Musique} & \textbf{Bamboogle} & \textbf{NQ} & \textbf{TriviaQA} & \textbf{PopQA} & \textbf{Avg.} \\ \midrule
\rowcolor{mylightblue} \text{Agent}$_2$ + \ours{}$_2$
& \textbf{45.82} & \textbf{45.30} & \textbf{20.27} & \textbf{48.00} & \textbf{42.43} & \textbf{69.02} & 44.14 & \textbf{45.00} \\
$\bullet$~Agent$_2$ + \ours{}$_1$
& 44.88 & 44.31 & 19.23 & 44.80 & 42.16 & 68.59 & \textbf{44.99} & 44.14 \\
$\bullet$~Agent$_1$ + \ours{}$_1$
& 40.44 & 44.49 & 16.63 & 44.00 & 39.69 & 65.80 & 44.75 & 42.26 \\
\quad$\bullet$~w/o GAC
& 38.91 & 43.42 & 16.71 & 43.20 & 39.19 & 63.91 & 42.66 & 41.14 \\
\quad$\bullet$~w/o LR
& 38.73 & 41.34 & 15.80 & 40.00 & 40.36 & 64.12 & 43.86 & 40.60 \\
\quad$\bullet$~w/o Question
& 39.52 & 44.02 & 16.13 & 43.20 & 39.30 & 64.96 & 43.89 & 41.57 \\
\bottomrule
\end{tabular}}
\caption{Ablation studies over key components.}
\label{tab:ablation}
\vspace{1mm}
\end{table*}

\subsection{Ablation Study} \label{sec:ablation}
We conduct ablation studies to examine the impact of two key components in our framework: (1) agent--retriever iterative optimization and (2) passage utility modeling.
Results are summarized in Table~\ref{tab:ablation}.
Here, ``$\text{Agent}_2$ + \ours{}$_2$'' denotes our final agent and retriever trained after two iterations.

\paragraph{Effect of Iterative Optimization.}
We first analyze the effect of iterative training on both the retriever and the search agent.
Replacing \ours{}$_2$ with the retriever \ours{}$_1$ obtained after the first iteration leads to an average performance drop of about 0.9 points.
This indicates that the second iteration further improves the retriever by leveraging higher-quality agent-generated queries.
Similar trends are also observed when using R1-Searcher (see Table~\ref{tab:ablation_r1-searcher}). We further replace $\text{Agent}_2$ with the agent trained after the first iteration (\ie, $\text{Agent}_1$), which results in an additional drop of about 1.9 points.
This suggests that compared with E5, optimized \ours{}$_1$ can better improve the RL training of the search agent.

\paragraph{Effect of Passage Utility Modeling.}
We next study the effectiveness of different utility signals, global answer correctness (GAC) and local relevance (LR), based on combination ``Agent$_1$ + \ours{}$_1$''.
Removing GAC (``w/o GAC'') or LR (``w/o LR'') from the utility modeling causes clear performance degradation, with average drops of about 1.1 and 1.7 points, respectively.
This confirms that both signals are essential for identifying positive passages for retriever training.
The larger performance drop of ``w/o LR'' indicates that local relevance plays a more important role in measuring passage utility. We also ablate the use of the original question in the retriever input (``w/o Question'').
The 0.7-points performance drop suggests that the original question helps the retriever better assess whether a passage could contribute to answering the question and generating the correct final answer.

\begin{figure}[!t]
	\centering
	\includegraphics[width=1\linewidth]{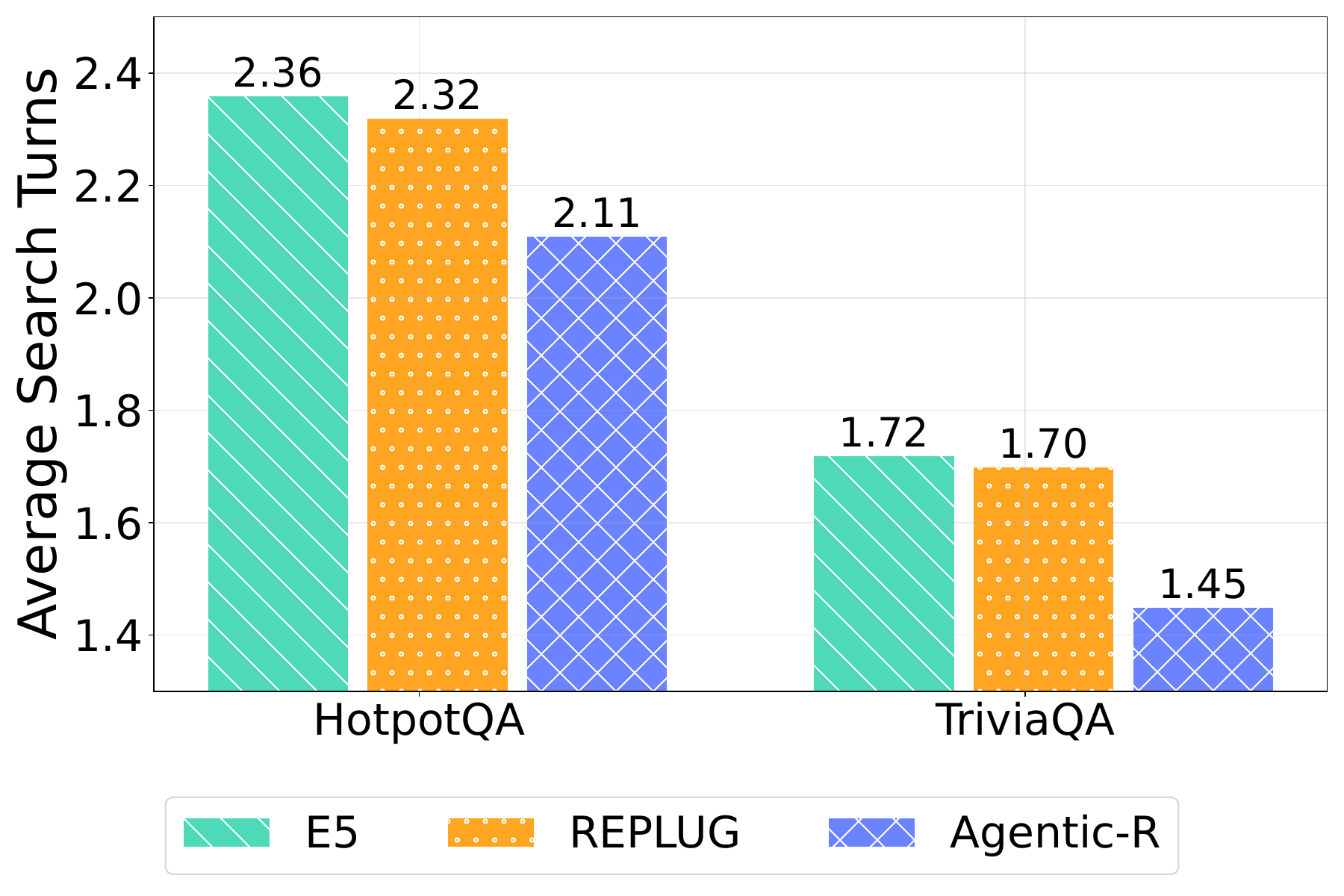}
	\caption{The comparison of search turns on different retrievers.}
	\label{fig:search_turns}
\end{figure}

\subsection{Search Turns Analysis}
\label{sec:search_turns}
In this section, we analyze the number of search turns taken by the agent to examine whether our \ours{} can also make the agent generate the answer with fewer search queries.
We conduct experiments on HotpotQA and TriviaQA using our search agent and \ours{} trained after two iterations, and compare against two baselines, E5 and REPLUG.
The results are shown in Figure~\ref{fig:search_turns}. 
From the results, \ours{} consistently reduces the number of search turns compared to E5 and REPLUG on both datasets.
For example, \ours{} reduces the average search turns by approximately 10\% on HotpotQA and 15\% on TriviaQA compared to REPLUG. These results indicate that \ours{} enables the agent to acquire more useful information per retrieval, allowing it to solve questions with fewer search turns.

\begin{figure}[!t]
	\centering
	\includegraphics[width=1\linewidth]{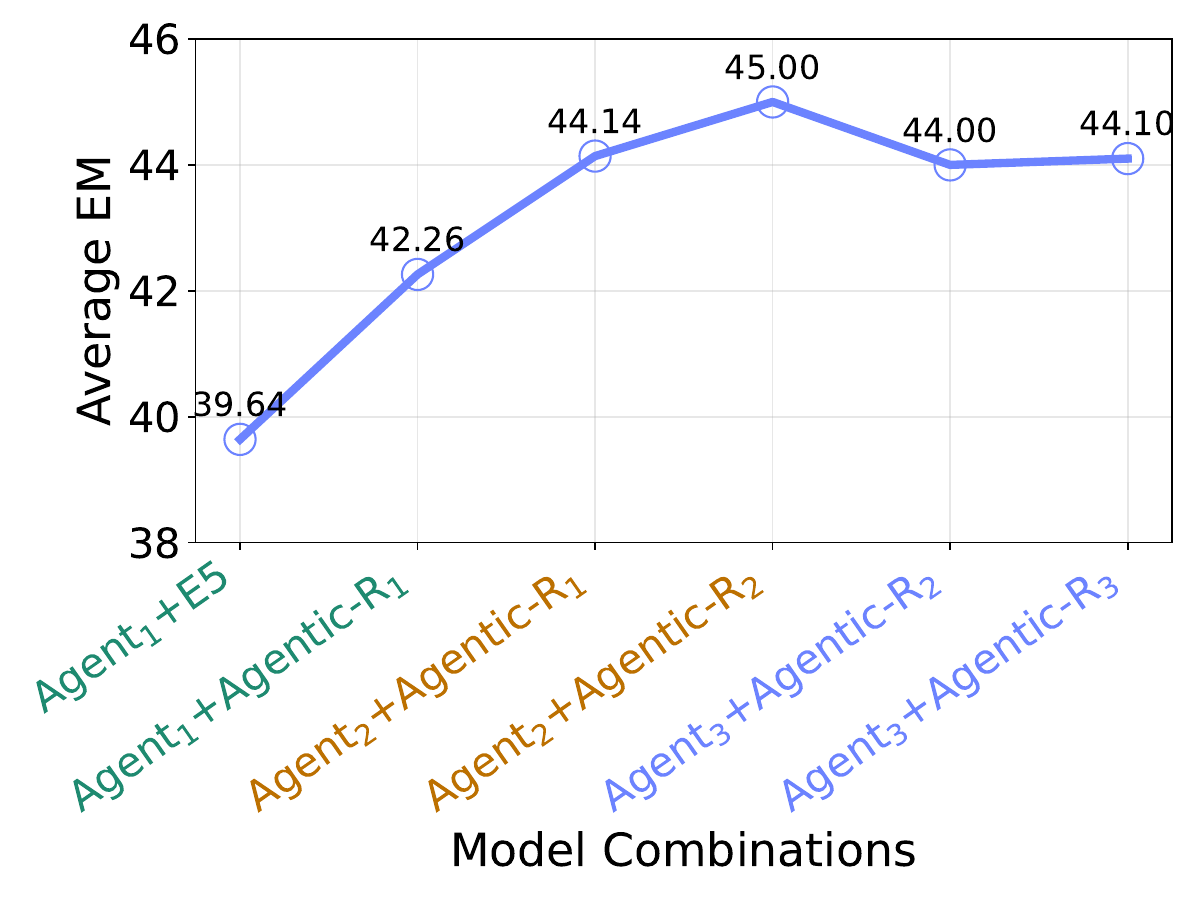}
    \caption{Effect of different iteration numbers ($K$) in agent--retriever optimization. Different colors on the x-axis indicate different iteration number.}
        	\label{fig:iteration_analysis}
\end{figure}

\subsection{Different Iteration Number $K$}
\label{sec:iteration_number}
In our main experiments, we set the round $K$ of agent--retriever iterative optimization as 2.
To examine whether the iterative training process converges, we further perform a third iteration and compare the performance after each round. We report the average EM score of all 7 datasets used in Table~\ref{tab:main_results} and show the results in Figure~\ref{fig:iteration_analysis}.

From the results, we observe consistent performance improvements during the first two iterations.
Specifically, \text{Agent}$_1$ + \ours{}$_1$ outperforms \text{Agent}$_1$ + E5 by approximately 2.6 points, and \text{Agent}$_2$ + \ours{}$_1$ further improves over \text{Agent}$_1$ + \ours{}$_1$ by about 1.9 points.
However, no further improvement is observed in the third iteration, where performance slightly degrades.
This indicates that the iterative optimization converges after two iterations, and additional iterations do not provide further benefits.

\section{Conclusion}
In this work, we propose a retriever training framework specifically designed for agentic search.
We first introduce a passage utility modeling strategy that captures both local relevance and global answer correctness, providing effective supervision for retriever training for multi-turn agentic search.
Furthermore, we develop an iterative agent--retriever training framework that continuously enhances the retriever's ability based on higher-quality agent-generated queries.
We conduct extensive experiments on multi-hop and single-hop QA benchmarks. Results demonstrate the superior performance of our retriever \ours{}. Further analyses show that \ours{} also reduces the number of search turns required by the agent, improving the efficiency of agentic search.

\section*{Limitations} \label{limitation}
Despite the strong empirical results, this work has several limitations.

First, our evaluation is conducted on widely used single-hop and multi-hop question answering benchmarks.
While these datasets cover a range of reasoning difficulties, they do not fully represent more challenging settings that require deeper or more abstract reasoning, such as expert-level or scientific reasoning tasks such as GPQA~\cite{GPQA}.
Future work could extend our framework to broader tasks.

Second, due to computational and memory constraints, we conduct our experiments with moderately sized search agents and retriever backbones.
Although our results across different backbones suggest favorable scaling trends, validating the effectiveness of our \ours in larger agents and retriever backbones remains an important direction for future work.



\clearpage
\appendix

\section{Baselines}
\label{app:baselines}

We compare our approach with two categories of retriever baselines: general-purpose embedding models and RAG-specific retrievers.

\paragraph{General-purpose embedding models.}
These retrievers are trained as universal text encoders and are widely adopted as off-the-shelf components in agentic search systems.
We include \textbf{E5}~\cite{e5} and \textbf{BGE}~\cite{bge} as representative baselines.
For E5, we use the \texttt{e5-base-v2}, and for BGE, we adopt \texttt{bge-base-en-v1.5}, both of which demonstrate strong performance on standard retrieval benchmarks.

\paragraph{RAG-specific retrievers.}
These methods train retrievers using generation feedback in single-turn RAG settings, aiming to capture passage utility beyond semantic similarity.

\textbf{LLM-Embedder}~\cite{llm-embedder} trains a unified embedding model with supervision from LLM-generated feedback to support retrieval-augmented generation across tasks.
We use the publicly available checkpoint\footnote{\url{https://huggingface.co/BAAI/llm-embedder}}.

\textbf{SCARLet}~\cite{SCARLet} estimates passage utility via perturbation-based attribution, explicitly modeling inter-passage interactions to better reflect a passage’s contribution to downstream generation.
We use the authors’ publicly released checkpoint for all experiments.

\textbf{REPLUG}~\cite{replug} defines passage utility by measuring how well each passage supports generating the ground-truth answer, using answer likelihood as the supervision signal.
As the official checkpoint is not released, we re-implement REPLUG and train it on the same datasets (HotpotQA and TriviaQA) with the same generator backbone, \texttt{Qwen2.5-7B-Base}, to ensure fair comparison.

\section{Implementation Details}
\label{app:implementation}
In this part, we will introduce the implementation details of our paper. 
\paragraph{Retriever Training.}
We initialize our retriever from e5-base-v2 and train it using contrastive learning with agent-generated supervision.
Training data are constructed following the passage utility modeling procedure described in Section~\ref{sec:utility_modeling}.
For each query, we use $N=16$ training passages in total, including one positive passage and multiple negatives.

We adopt mean pooling over token embeddings and apply $\ell_2$ normalization.
The retriever is trained for 2 epochs with a learning rate of $2 \times 10^{-5}$ and a per-device batch size of 32.
We use a temperature of 0.01 in the contrastive loss and enable in-batch and cross-device negatives, resulting in a large and diverse negative set.
The maximum input length is set to 512 tokens for both queries and passages.
All retriever models are trained on the December 2018 Wikipedia dump, which serves as the retrieval corpus.

\paragraph{Agent Training.} \label{app:agent_training}
We train the search agent following the same reinforcement learning approach of Search-R1~\cite{search-r1}. Specifically, the policy LLM is initialized from Qwen2.5-7B-Base, and the agent is optimized using Proximal Policy Optimization (PPO)~\cite{ppo}. Unlike standard PPO settings, where all tokens are generated by the policy model, the rollout sequence in agentic search contains both LLM-generated tokens and retrieved tokens from external passages.
We therefore apply token-level loss masking to ensure that policy optimization is performed only on LLM-generated tokens, while retrieved tokens are excluded from gradient updates. This design allows the agent to learn effective reasoning and search behaviors while preventing spurious updates on retrieved tokens, resulting in more stable training for search-augmented generation. Refer to Search-R1~\cite{search-r1} for more details.

During RL training, the learning rate is set to $1 \times 10^{-6}$ for the policy model and $1 \times 10^{-5}$ for the value model.
We train the agent for 500 optimization steps, using Generalized Advantage Estimation (GAE) with $\lambda=1$ and $\gamma=1$.
The total batch size is 512, with a mini-batch size of 256 and a micro-batch size of 64.
The maximum sequence length is 4,096 tokens, including up to 500 tokens for the generated response and up to 500 tokens for retrieved passages.

To reduce memory consumption, we enable gradient checkpointing and employ Fully Sharded Data Parallel (FSDP) with CPU offloading.
Model checkpoints are saved every 50 steps.
If training instability is observed, we select the most recent stable checkpoint based on the reward curve; otherwise, the final checkpoint is used for evaluation.

\begin{table*}[!t]
\small
\centering
\setlength{\tabcolsep}{2.8mm}{
\begin{tabular}{lcccccccc}
\toprule
\textbf{} & \multicolumn{4}{c}{\textbf{Multi-Hop QA}} & \multicolumn{3}{c}{\textbf{General QA}} & \textbf{} \\
\cmidrule(lr){2-5} \cmidrule(lr){6-8}
\textbf{Methods} & \textbf{HotpotQA} & \textbf{2Wiki} & \textbf{Musique} & \textbf{Bamboogle} & \textbf{NQ} & \textbf{TriviaQA} & \textbf{PopQA} & \textbf{Avg.} \\ \midrule
\rowcolor{mygray}
\multicolumn{9}{c}{\textbf{Our Search Agent (in-domain)}} \\ \midrule
LLM-Embedder & 35.92 & 39.79 & 14.97 & 38.40 & 39.58 & 59.01 & 40.93 & 38.37 \\
BGE-base & 37.51 & 39.86 & 14.27 & 40.00 & 37.39 & 59.94 & 39.53 & 38.36 \\
SCARLet & 37.82 & 39.71 & 13.98 & 39.20 & 37.31 & 59.98 & 39.55 & 38.22 \\
E5-base & 36.32 & 42.04 & 16.13 & 36.80 & 39.77 & 64.35 & 42.09 & 39.64 \\
REPLUG & 38.05 & 40.78 & 16.38 & 37.60 & 39.08 & 63.00 & 39.72 & 39.23 \\
E5-large & 38.69 & 40.53 & {\ul 16.88} & \textbf{44.00} & \textbf{41.10} & 63.49 & 40.98 & 40.81 \\
\rowcolor{mylightblue} \ours{$_\text{E5-base}$} & {\ul 40.44} & \textbf{44.49} & 16.63 & \textbf{44.00} & 39.69 & {\ul 65.80} & \textbf{44.75} & {\ul 42.26} \\
\rowcolor{mylightblue} \ours{$_\text{BGE-base}$} & 40.28 & 43.82 & 15.97 & {\ul 42.40} & 38.39 & 64.00 & 43.30 & 41.17 \\
\rowcolor{mylightblue} \ours{$_\text{E5-large}$} & \textbf{41.49} & {\ul 44.40} & \textbf{18.08} & \textbf{44.00} & {\ul 39.97} & \textbf{66.26} & {\ul 44.40} & \textbf{42.66} \\ \midrule
\rowcolor{mygray}
\multicolumn{9}{c}{\textbf{SimpleDeepSearcher (out-of-domain)}} \\ \midrule
LLM-Embedder & 35.01 & 32.52 & 13.90 & 40.80 & 33.62 & 59.59 & 37.72 & 36.17 \\
BGE-base & 37.31 & 33.95 & 14.06 & 42.40 & 32.82 & 60.73 & 36.83 & 36.87 \\
SCARLet & 37.39 & 33.44 & 14.39 & 35.19 & 32.38 & 60.81 & 36.93 & 35.79 \\
E5-base & 36.61 & 34.01 & 15.55 & 40.80 & 33.65 & 62.48 & 37.09 & 37.17 \\
REPLUG & 37.39 & 34.08 & {\ul 15.80} & 41.60 & 33.71 & 63.59 & 37.68 & 37.69 \\
E5-large & 37.59 & 33.20 & 15.72 & 43.20 & \textbf{34.90} & 63.58 & 37.63 & 37.97 \\
\rowcolor{mylightblue} \ours{$_\text{E5-base}$} & {\ul 39.98} & {\ul 37.43} & 15.18 & {\ul 44.80} & 33.98 & {\ul 65.16} & \textbf{39.46} & {\ul 39.43} \\
\rowcolor{mylightblue} \ours{$_\text{BGE-base}$} & 38.83 & 35.92 & 15.26 & 40.80 & 33.21 & 64.31 & 38.38 & 38.10 \\
\rowcolor{mylightblue} \ours{$_\text{E5-large}$} & \textbf{40.25} & \textbf{38.12} & \textbf{16.09} & \textbf{46.40} & {\ul 34.65} & \textbf{65.58} & {\ul 39.27} & \textbf{40.05} \\ \bottomrule
\end{tabular}}
\caption{The performance of \ours{} using different backbones. Our search agent and \ours{} are trained for one iteration. }
\label{tab:different_backbone}
\end{table*}

\begin{table*}[!t]
\small
\centering
\setlength{\tabcolsep}{2.1mm}{
\begin{tabular}{lcccccccc}
\toprule
\textbf{} & \multicolumn{4}{c}{\textbf{Multi-Hop QA}} & \multicolumn{3}{c}{\textbf{General QA}} & \textbf{} \\
\cmidrule(lr){2-5} \cmidrule(lr){6-8}
\textbf{Methods} & \textbf{HotpotQA} & \textbf{2Wiki} & \textbf{Musique} & \textbf{Bamboogle} & \textbf{NQ} & \textbf{TriviaQA} & \textbf{PopQA} & \textbf{Avg.} \\ \midrule
R1-Searcher + \ours{}$_2$ & \textbf{47.68} & \textbf{49.07} & \textbf{22.54} & \textbf{41.60} & 39.63 & \textbf{62.52} & 42.43 & \textbf{43.64} \\
R1-Searcher + \ours{}$_1$ & 46.86 & 48.64 & 21.39 & 35.19 & \textbf{39.94} & 62.33 & \textbf{43.07} & 42.49 \\ \bottomrule
\end{tabular}}
\caption{The performance comparison between \ours{}$_1$ and \ours{}$_2$ based on R1-Searcher.}
\label{tab:ablation_r1-searcher}
\end{table*}

\paragraph{Iterative Training Setup.}
During iterative agent--retriever optimization, we alternate between training the search agent and updating the retriever.
At each iteration, the retriever is fixed while training the agent, and the trained agent is then used to generate new trajectories for retriever training.
We perform two iterations in total, as additional iterations do not yield further improvements (see Section~\ref{sec:ablation}). All experiments are conducted on a single node with 8*A800 80G GPUs.

\section{Additional Experiments}

\subsection{Different Backbone of \ours{}}
\label{app:backbone}

During our previous experiments, we used E5-base as the backbone of \ours{}.
In this section, we evaluate whether our retriever training framework generalizes to other backbone models by training \ours{} on BGE-base\footnote{\url{https://huggingface.co/BAAI/bge-base-en-v1.5}} and E5-large\footnote{\url{https://huggingface.co/intfloat/e5-large-v2}}.
For simplicity, both the search agent and the retriever are trained for a single iteration in this experiment.

As shown in Table~\ref{tab:different_backbone}, \ours{} consistently outperforms all baseline retrievers across all tested backbones and search agents, demonstrating that our method generalizes well beyond a specific embedding model.
Moreover, \ours{} yields substantial improvements over its corresponding backbone retriever; for example, under our search agent, \ours{}$_{\text{BGE-base}}$ improves upon BGE-base by approximately 2.8 average EM points.
We further observe a clear scaling trend with respect to retriever capacity: the larger backbone E5-large consistently outperforms E5-base.
A similar trend is also observed between \ours{}$_{\text{E5-base}}$ and \ours{}$_{\text{E5-large}}$.

\begin{table*}[!t]
\small
\centering
\setlength{\tabcolsep}{2.1mm}{
\begin{tabular}{lcccccccc}
\toprule
\textbf{} & \multicolumn{4}{c}{\textbf{Multi-Hop QA}} & \multicolumn{3}{c}{\textbf{General QA}} & \textbf{} \\
\cmidrule(lr){2-5} \cmidrule(lr){6-8}
\textbf{Methods} & \textbf{HotpotQA} & \textbf{2Wiki} & \textbf{Musique} & \textbf{Bamboogle} & \textbf{NQ} & \textbf{TriviaQA} & \textbf{PopQA} & \textbf{Avg.} \\ \midrule
\multicolumn{9}{c}{\textbf{Our Search Agent}} \\ \midrule
E5 & 36.32 & 42.04 & 16.13 & 36.80 & {\ul 39.77} & 64.35 & 42.09 & 39.64 \\
REPLUG & 38.05 & 40.78 & 16.38 & 37.60 & 39.08 & 63.00 & 39.72 & 39.23 \\
\ours{} & {\ul 40.44} & {\ul 44.49} & {\ul 16.63} & \textbf{44.00} & 39.69 & \textbf{65.80} & {\ul 44.75} & \textbf{42.26} \\
\ours{} (w/ historical queries) & \textbf{40.47} & \textbf{44.65} & \textbf{17.04} & {\ul 38.40} & \textbf{40.02} & {\ul 65.67} & \textbf{44.87} & {\ul 41.59} \\ \midrule
\multicolumn{9}{c}{\textbf{R1-Searcher}} \\ \midrule
E5 & {\ul 43.56} & {\ul 46.33} & \textbf{21.39} & \textbf{44.00} & 39.39 & 58.69 & 38.31 & {\ul 41.67} \\
REPLUG & 40.68 & 39.66 & {\ul 18.32} & {\ul 41.60} & \textbf{40.94} & \textbf{62.39} & 42.11 & 40.81 \\
\ours & \textbf{46.86} & \textbf{48.64} & \textbf{21.39} & 35.19 & {\ul 39.94} & {\ul 62.33} & {\ul 43.07} & \textbf{42.49} \\
\ours~(w/ historical queries) & 43.13 & 42.95 & 15.30 & 31.20 & 39.63 & 62.18 & \textbf{43.17} & 39.65 \\ \bottomrule
\end{tabular}}
\caption{The performance of \ours{} when using historical queries as additional training input.}
\label{tab:historical_queries}
\end{table*}

\subsection{Training Input}
\label{app:input_analysis}

In the main experiments, we construct the retriever input using only the original question $Q$ and the current-turn query $q_i$.
In this part, we further investigate whether incorporating historical queries from previous turns can benefit retriever training.

Specifically, we concatenate the original question $Q$, all historical queries $\{q_1, \ldots, q_{i-1}\}$, and the current query $q_i$ with separator tokens as the retriever input:
\begin{equation}
\label{equ:query_input_history}
x_i = Q \; [\text{SEP}] \; q_1 \; [\text{SEP}] \; \cdots q_{i-1} \; [\text{SEP}] \; q_i .
\end{equation}
The same input format is also used at inference time.
We conduct this ablation using a single iteration of agent--retriever training and compare it with the default setting that excludes historical queries.

The results are shown in Table~\ref{tab:historical_queries}.
We observe that incorporating historical queries (\ie, \ours~(w/ historical queries)) degrades average performance for \ours{} under both the in-domain search agent and the out-of-domain R1-Searcher.
The performance drop is particularly obvious under R1-Searcher, where \ours{} with historical queries even underperforms general-purpose retrievers such as E5 and REPLUG.

This behavior can be attributed to the nature of agentic search queries.
In agentic search, the current query $q_i$ is typically generated with a clear and explicit intent, and does not involve anaphoric references or contextual ambiguity.
As a result, historical queries provide limited additional information, while introducing extra noise that may distract the retriever from the core intent of the current query.
These results further justify our design choice of using only $(Q, q_i)$ as the retriever input in \ours{}.

\begin{table*}[!t]
\centering
\caption{Case study comparing E5 and \ours{} on a multi-hop query of HotpotQA. The \textcolor{red}{red-highlighted} text contains irrelevant information that misled the generation of the correct answer, while the \textcolor{green}{green-highlighted} text represents key evidence that contributed to generating the correct answer.}
\fontsize{9pt}{11pt}\selectfont
\begin{tabular}{p{0.98\linewidth}}
\toprule

\rowcolor{gray!20}\textbf{Query} \\
Who founded the honky tonk that is at the center of John Travolta's third major acting role? \\
\midrule

\rowcolor{gray!20}\textbf{Gold Answer} \\
Mickey Gilley \\
\midrule

\rowcolor{gray!20}\textbf{E5 (Failed)} \\
\midrule

\textcolor{deepblue}{\texttt{<think>}}
John Travolta is an actor, so I will search for his filmography to find his third major acting role.
\textcolor{deepblue}{\texttt{</think>}}

\textcolor{blue}{\texttt{<search>}}
John Travolta third major acting role 
\textcolor{blue}{\texttt{</search>}}

\textcolor{orange}{\texttt{<information>}} \\
Doc 1(Title: \"John Travolta\") John Travolta first became known in the 1970s, after appearing on the television series
Welcome Back, Kotter (1975--1979) and starring in the box office successes
Saturday Night Fever (1977) and Grease (1978).
His acting career declined through the 1980s, but enjoyed a resurgence in the 1990s with
his role in Pulp Fiction (1994), and he has since starred in films such as
\textcolor{red}{Get Shorty, Broken Arrow, Face/Off, Swordfish, Be Cool, Wild Hogs, Hairspray}, ... \\
\textcolor{orange}{\texttt{</information>}}
\\ \midrule

\textcolor{deepblue}{\texttt{<think>}}
\textcolor{red}{Assuming the movie is Get Shorty}, I now search for the honky tonk.
\textcolor{deepblue}{\texttt{</think>}}

\textcolor{blue}{\texttt{<search>}}
Get Shorty honky tonk
\textcolor{blue}{\texttt{</search>}}

\textcolor{orange}{\texttt{<information>}} \\
Doc1: ... Doc2 (Honky-tonk): A honky-tonk (also called honkatonk, honkey-tonk, or tonk) is both a bar that provides
country music for the entertainment of its patrons and the style of music played in
such establishments. Bars of this kind are common in the South and Southwest
United States... \\
\textcolor{orange}{\texttt{</information>}}
\\ \midrule
\textcolor{deepblue}{\texttt{<think>}}
Search for the founder of the assumed honky tonk.
\textcolor{deepblue}{\texttt{</think>}}

\textcolor{blue}{\texttt{<search>}}
The Big Enchilada honky tonk founder
\textcolor{blue}{\texttt{</search>}}

\textcolor{orange}{\texttt{<information>}} \\
Doc 1(Title: Honky-tonk) Honky-tonk: A honky-tonk (also called honkatonk, honkey-tonk, or tonk) is both a bar that provides country music for the entertainment of its patrons and the style of music played in such establishments...\\
\textcolor{orange}{\texttt{</information>}}
\\ \midrule
\textcolor{purple}{\texttt{<answer>}}  
\textcolor{red}{Eddie Wilson}  
\textcolor{purple}{\texttt{</answer>}} \\

\midrule

\rowcolor{gray!20}\textbf{\ours{} (Successful)} \\
\midrule

\textcolor{deepblue}{\texttt{<think>}}
John Travolta is an actor, so I will search for his filmography to find his third major acting role.
\textcolor{deepblue}{\texttt{</think>}}

\textcolor{blue}{\texttt{<search>}}
John Travolta third major acting role
\textcolor{blue}{\texttt{</search>}}

\textcolor{orange}{\texttt{<information>}} \\
Doc1 (Urban Cowboy): \textcolor{green}{Urban Cowboy} is a 1980 American romantic drama film about the love-hate relationship between Buford Uan ``Bud'' Davis (John Travolta) and Sissy (Debra Winger). It was \textcolor{green}{John Travolta's third major film role} after Saturday Night Fever and Grease. Much of the action centers around activities at \textcolor{green}{Gilley's Club}, a football-field-sized honky tonk in Pasadena, Texas... \\
\textcolor{orange}{\texttt{</information>}}
\\ \midrule
\textcolor{deepblue}{\texttt{<think>}}
Now that I have the name of the honky tonk, I will search for information about Gilley's Club and its founder.
\textcolor{deepblue}{\texttt{</think>}}

\textcolor{blue}{\texttt{<search>}}
Gilley's Club founder
\textcolor{blue}{\texttt{</search>}}

\textcolor{orange}{\texttt{<information>}} \\
Doc2 (Gilley's Club): Gilley's was a bar/honky tonk founded in 1971 by country singer \textcolor{green}{Mickey Gilley} in
Pasadena, Texas. It was the central location in the 1980 movie Urban Cowboy. It was a huge building with a corrugated steel roof that housed multiple bars and mechanical bulls. Connected to the club was a small rodeo arena that would also host both bicycle and motorcycle motocross races... \\
\textcolor{orange}{\texttt{</information>}}
\\ \midrule
\textcolor{purple}{\texttt{<answer>}}  
\textcolor{green}{Mickey Gilley}  
\textcolor{purple}{\texttt{</answer>}} \\

\bottomrule
\end{tabular}
\label{tab:case}
\end{table*}

\section{Case Study}
To better understand how \ours{} improves the performance of the search agent, we conduct a case study comparing the trajectories produced by E5 and \ours{} based on our trained agent. Note that both \ours{} and our trained agent are trained for 2 iterations. The result is shown in Table~\ref{tab:case}.

\section{Use of AI Assistants}
We use ChatGPT to improve the presentations of this paper.\footnote{\url{https://chatgpt.com/}}

\begin{figure*}[!tb]
  \centering
  \includegraphics[width=1\linewidth]{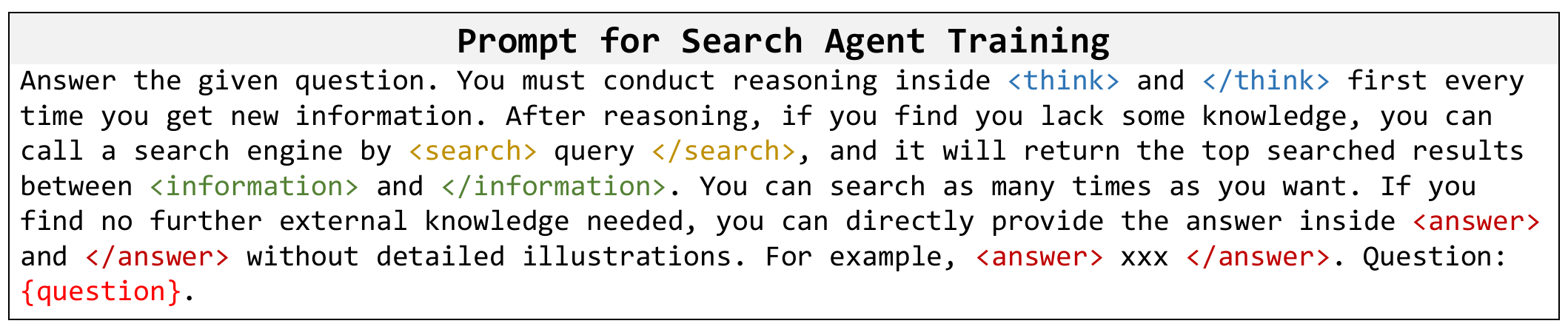}
  \caption{The input prompt for training our search agent.}
  \label{fig:searchr1_prompt}
\end{figure*}

\begin{figure*}[!tb]
  \centering
  \includegraphics[width=1\linewidth]{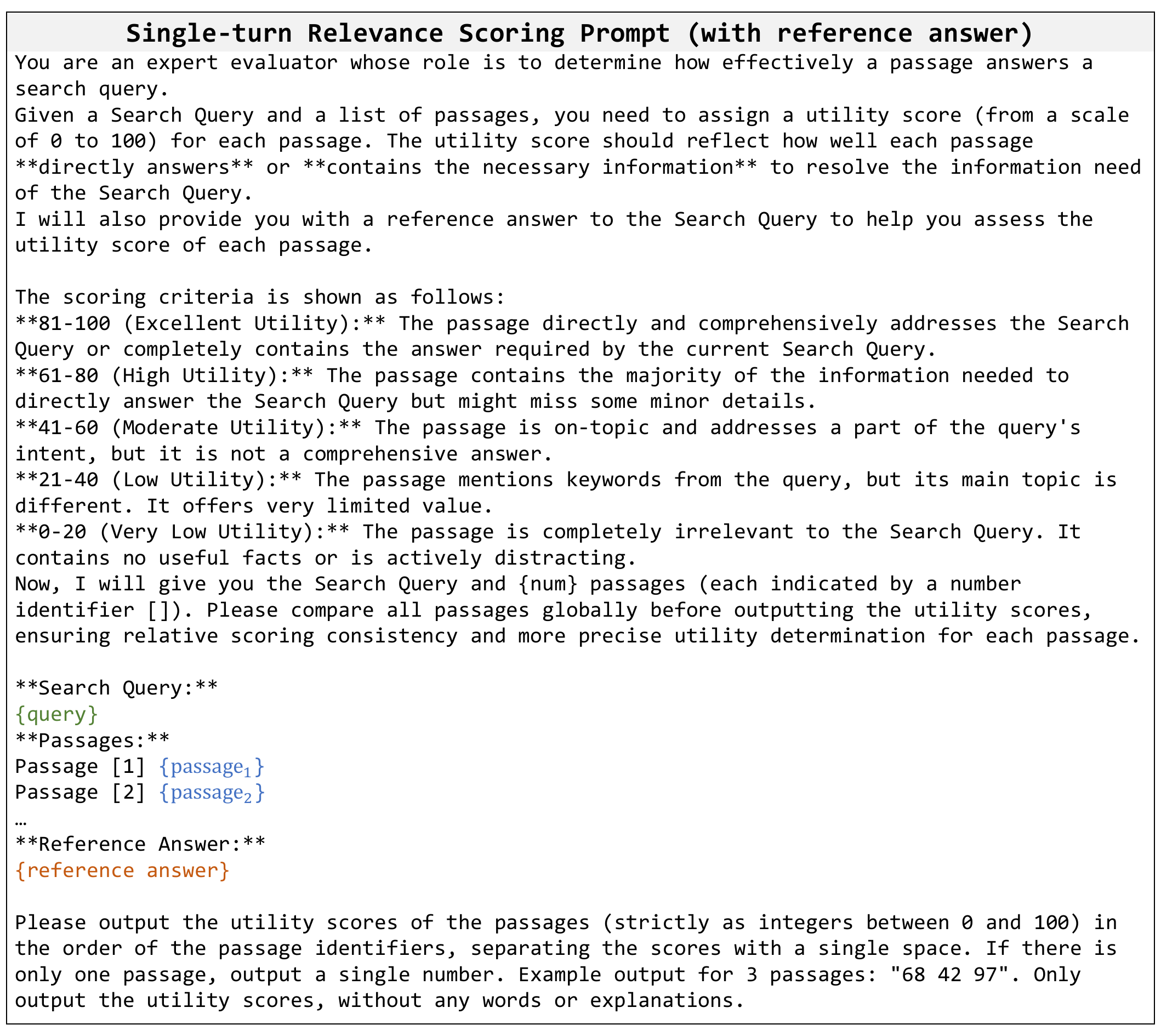}
  \caption{The prompt for single-turn relevance scoring with sub-answer.}
  \label{fig:scoring_with_answer}
\end{figure*}

\begin{figure*}[!tb]
  \centering
  \includegraphics[width=1\linewidth]{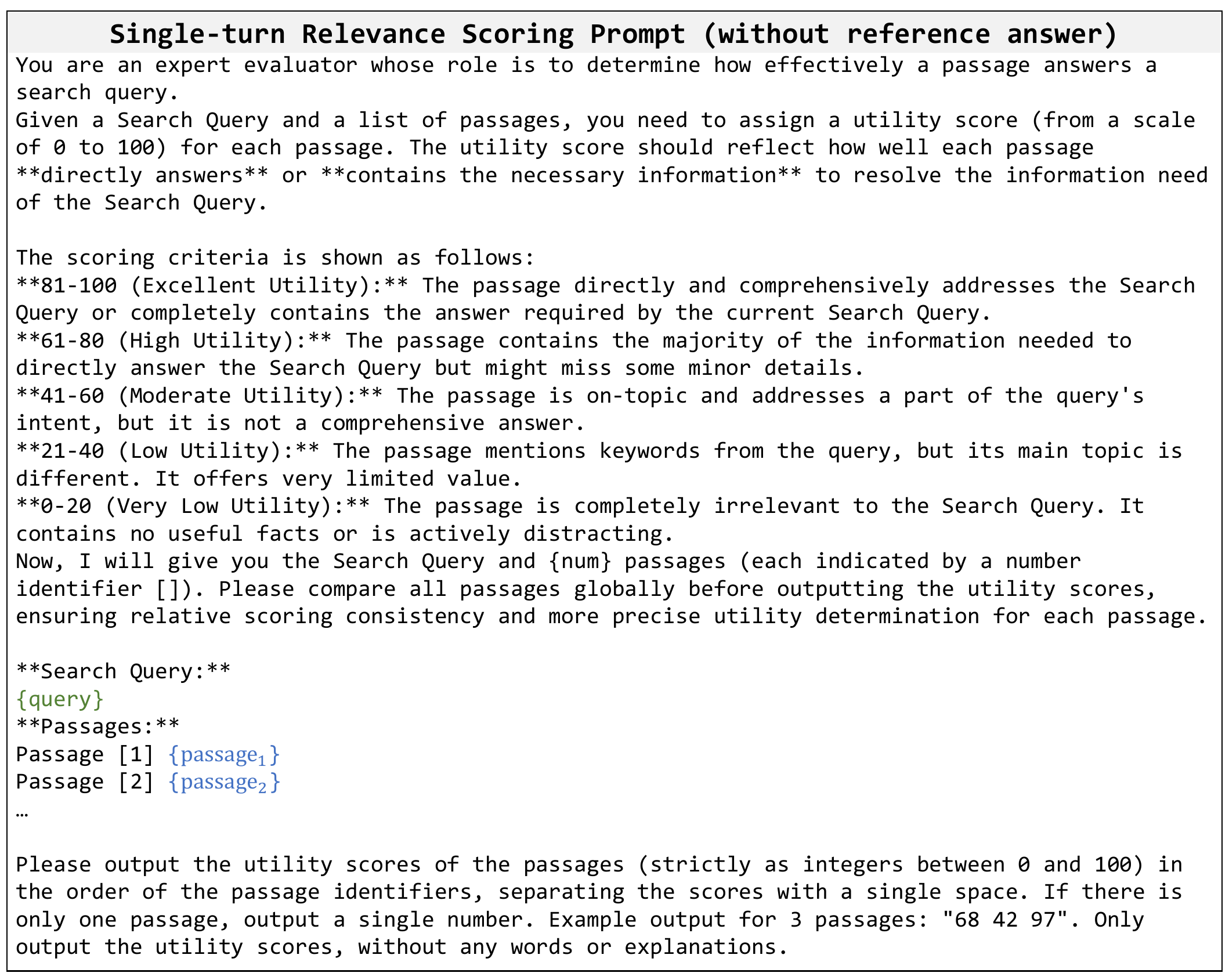}
  \caption{The prompt for single-turn relevance scoring without sub-answer.}
  \label{fig:scoring_without_answer}
\end{figure*}

\begin{figure*}[!tb]
  \centering
  \includegraphics[width=1\linewidth]{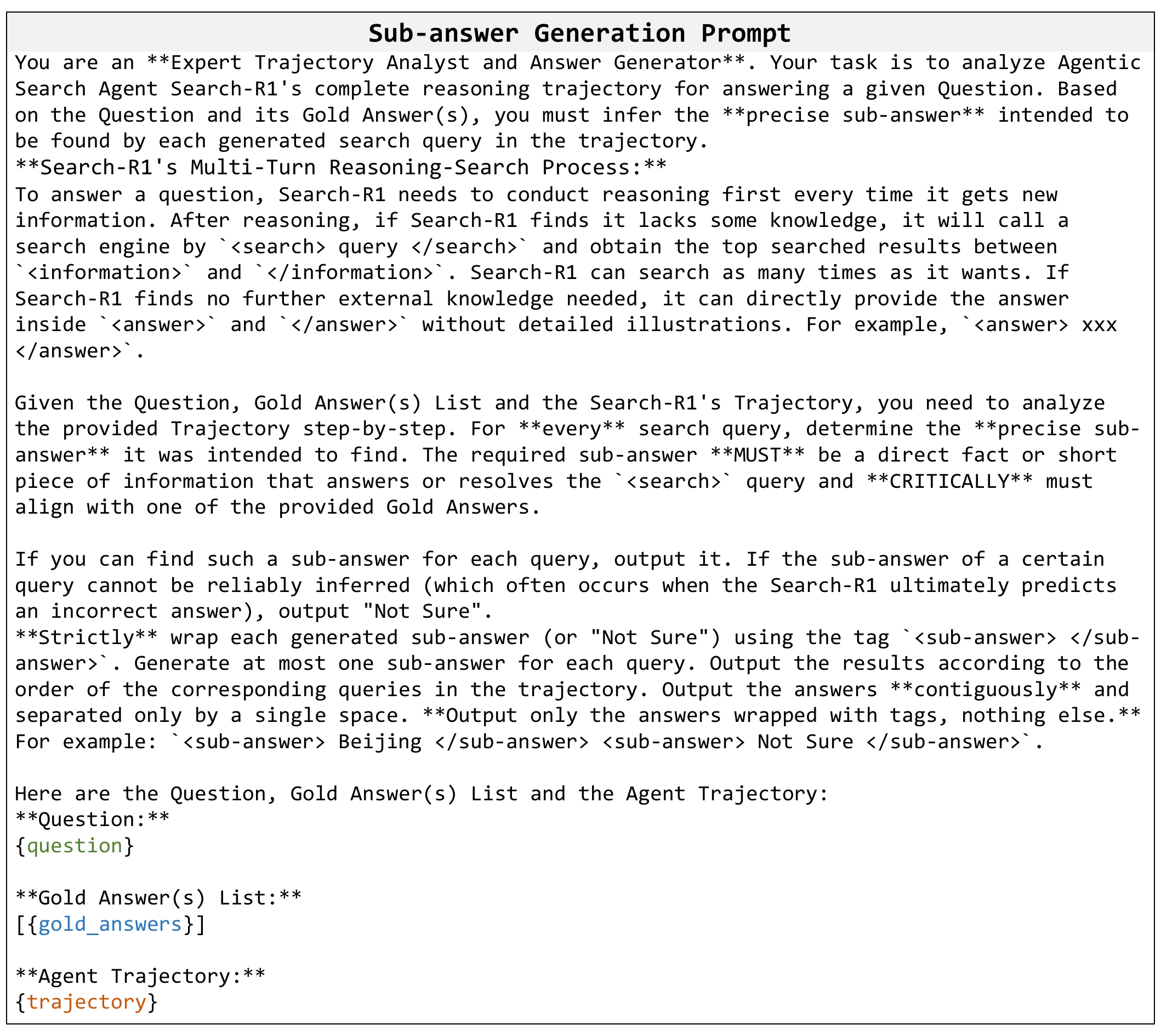}
  \caption{The prompt for generating sub-answers.}
  \label{fig:subanswer_generation}
\end{figure*}

\end{document}